\documentclass{Interspeech2024}
\usepackage{footnote}
\makesavenoteenv{tabular}
\makesavenoteenv{table}
\usepackage{makecell} 
\usepackage{multirow}
\usepackage{cite}

\interspeechcameraready

\title{MINT: Boosting Audio-Language Model via Multi-Target Pre-Training and Instruction Tuning}

\name[affiliation={1*}]{Hang}{Zhao}
\name[affiliation={2*}]{Yifei}{Xin}
\name[affiliation={1}]{Zhesong}{Yu}
\name[affiliation={1}]{Bilei}{Zhu}
\name[affiliation={1}]{Lu}{Lu}
\name[affiliation={1}]{Zejun}{Ma}

\address{$^{1}$Bytedance Inc. $^{2}$Peking University, China \thanks{$^{*}$ The first two authors have equal contribution.}}
\email{zhaohang.ai@bytedance.com, 2101212833@stu.pku.edu.cn}


\begin{document}
\maketitle
\begin{abstract}
In the realm of audio-language pre-training (ALP), the challenge of achieving cross-modal alignment is significant. Moreover, the integration of audio inputs with diverse distributions and task variations poses challenges in developing generic audio-language models. In this study, we present MINT, a novel ALP framework boosting audio-language models through multi-target pre-training and instruction tuning. MINT leverages the strength of frozen pre-trained audio encoders and large language models (LLM) to improve audio-language pre-training, enabling effective transferablility to both audio-text understanding and generation tasks. To address the modality gap, we introduce Bridge-Net, a trainable module that enhances cross-modality alignment and the model's ability to follow instructions for a variety of audio-text tasks. Bridge-Net is pivotal within MINT, initially enhancing audio-language representation learning through a multi-target pre-training approach. Subsequently, Bridge-Net further boosts audio-to-language generative learning by integrating a frozen language model with instruction tuning. This integration empowers MINT to extract features in a flexible and effective manner, specifically tailored to the provided instructions for diverse tasks. Experimental results demonstrate that MINT attains superior performance across various audio-language understanding and generation tasks, highlighting its robust generalization capabilities even in zero-shot scenarios.
\end{abstract}
\section{Introduction}
Recent advancements in audio-language pre-training (ALP) have markedly enhanced performance across a range of downstream tasks \cite{elizalde2023clap,wu2023large,deshmukh2023pengi}. In the field of natural language processing (NLP) \cite{kenton2019bert,cheng2023ml,cheng2023mrrl}, large-scale pre-training and instruction tuning have shown promise in developing versatile language models capable of solving diverse tasks. A growing interest has emerged in utilizing large language models (LLM) to further enrich ALP capabilities. However, narrowing the modality gap and devising audio-language models that can effectively follow instructions remain under-explored. In this work, we introduce MINT: an audio-language pre-training framework with multi-target leaning and instruction tuning to enhance ALP. MINT is designed to support various audio-related tasks, including close-ended tasks like classification \cite{deshmukh2023pengi,23d_interspeech} and retrieval \cite{2023pooling,2023cooperative}, as well as open-ended tasks such as audio captioning \cite{gontier2021automated,mei2021audio}. We leverage off-the-shelf frozen pre-trained audio encoders to extract effective audio representations and utilize frozen pre-trained large language models for powerful language generation and zero-shot transferability.

In order to effectively leverage pre-trained audio encoders and LLMs in ALP, it is crucial to facilitate cross-modal alignment. However, since they have not been exposed to each other during pre-training, freezing them poses a significant challenge in achieving audio-language alignment. To overcome this issue, we introduce Bridge-Net, which incorporates a set of learnable query vectors to extract audio features from the frozen audio encoder. Bridge-Net leverages Q-Former \cite{li2023blip} to align the output feature space of the frozen audio encoder with the language model, which not only reduces the computational cost of training but also effectively mitigates the issue of catastrophic forgetting. Acting as an information bottleneck between the frozen audio encoder and the frozen LLM, Bridge-Net selectively provides the most relevant audio features to the LLM for generating the desired output text. In the pre-training stage, we focus on audio-language representation learning with audio-text pairs, guiding Bridge-Net to extract audio representations that are most relevant to the corresponding text. In the instruction tuning stage, we transit the training objective to audio-to-language generative learning. We feed the output of Bridge-Net, after passing through a linear projection, to a frozen LLM along with text instructions.

To enhance the model's generalization to unseen data, we introduce an instruction-aware audio feature extraction mechanism during instruction tuning phase. This mechanism allows for flexible and informative feature extraction aligned with the provided instructions for various tasks. The system not only feeds textual instructions to the frozen LLM but also to the Bridge-Net, enabling it to extract instruction-specific audio features from the frozen audio encoder. By integrating instruction tuning, our approach achieves improved effectiveness and flexibility in addressing downstream tasks.
The principal advantages of MINT include:
\begin{itemize}
\item MINT presents an effective ALP framework of leveraging frozen pre-trained audio models and language models. In order to bridge the modality gap, MINT introduces Bridge-Net, which is trained using multiple objectives and tuned to follow instructions.
\item During the instruction tuning stage, MINT incorporates an instruction-aware audio feature extraction mechanism, which proves to be powerful for flexible and informative audio feature extraction in line with the given instructions. 
\item MINT significantly improves various audio understanding and audio-to-text generation tasks, thus establishing a strong baseline for general-purpose ALP.
\end{itemize}

\begin{figure}[t]
  \centering
  \includegraphics[width=1.0\linewidth]{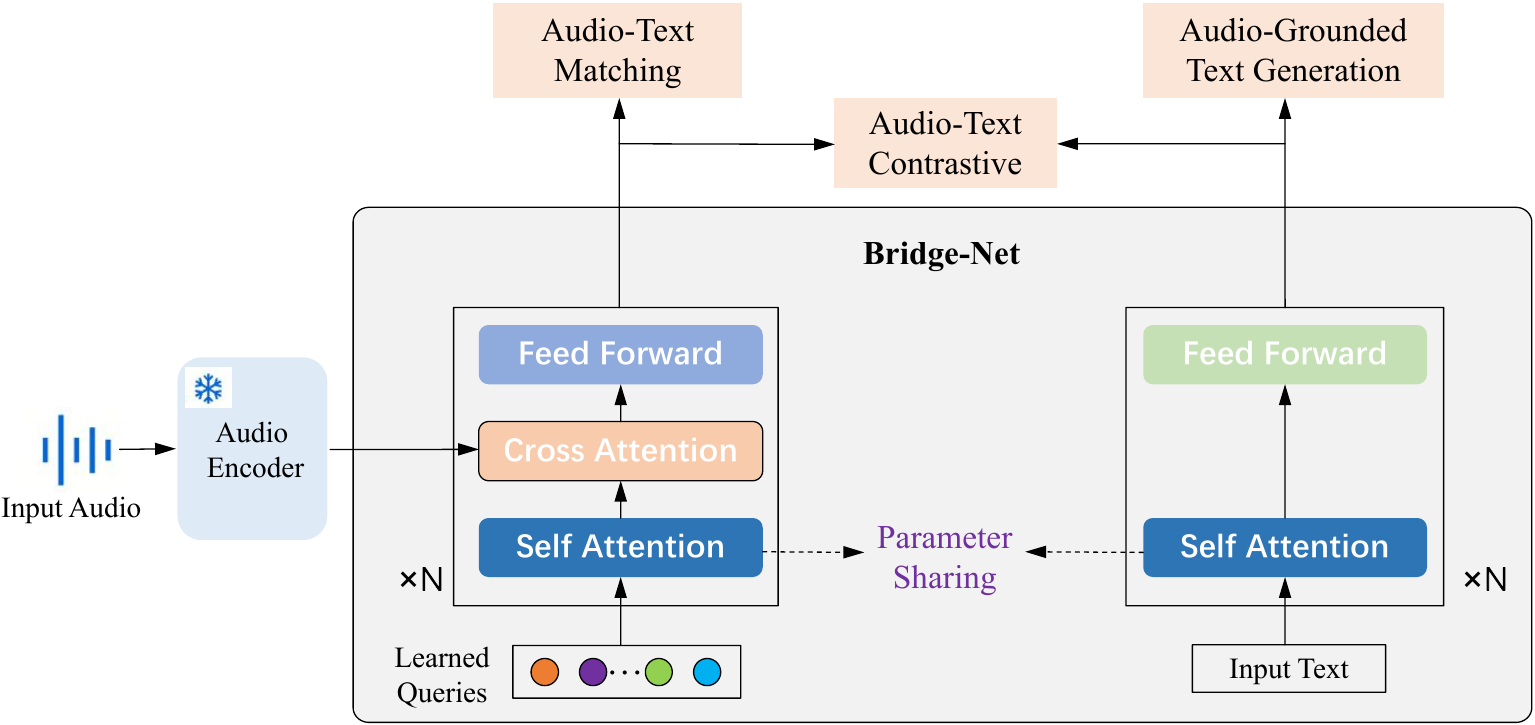}
  \caption{Model architecture of our MINT’s audio-language representation learning.}
  \vspace*{-\baselineskip}
\end{figure}
\section{Proposed methods}
\label{proposed method}
We introduce MINT, an advanced method that enhances audio-language models through multi-target learning and instruction tuning. MINT effectively utilizes frozen pre-trained models and introduces Bridge-Net to bridge the modality gap. In the pre-training stage, MINT performs audio-language representation learning with a frozen audio encoder. Bridge-Net is optimized with multi-target learning to learn audio representations that are most relevant to the corresponding texts. In the instruction tuning stage, MINT conducts an audio-to-language generative learning with a frozen LLM. Various audio tasks are framed as inputting both audio and text and outputting text, allowing the Bridge-Net to align instruction-aware audio features with the frozen LLM as a text generation condition. Bridge-Net is trained using the standard language modeling procedure to generate task responses. In this section, we provide a detailed explanation of the overall model architecture, followed by a comprehensive introduction to the training procedures.

\subsection{Model architecture}
We introduce Bridge-Net as a trainable module to align audio features to the text space, which uses Q-Former to narrow the gap between a frozen audio encoder and a frozen LLM. Bridge-Net consists of two transformer submodules that share the same self-attention layers: (1) an audio transformer that interacts with the frozen audio encoder to extract acoustic features, and (2) a text transformer that can operate as both a text encoder and a text decoder. Within Bridge-Net, a set of learnable query embeddings are provided as input. These queries interact with each other through self-attention layers, and interact with the frozen audio features through cross-attention layers. Additionally, the queries can interact with the text branch using the same self-attention layers. Q-Former is initialized with pre-trained weights from the BERT-base model \cite{kenton2019bert}, with the cross-attention layers being randomly initialized.

During the instruction tuning phase, we enhance the Bridge-Net to an instruction-aware architecture by incorporating the instruction text tokens as additional input. The instruction interacts with the query embeddings through the self-attention layers of Q-Former, prompting the extraction of audio features that are most relevant to the tasks. Consequently, the LLM receives acoustic features that are informative and conducive to following the given instruction.

\begin{figure}[t]
  \centering
  \includegraphics[width=1.0\linewidth]{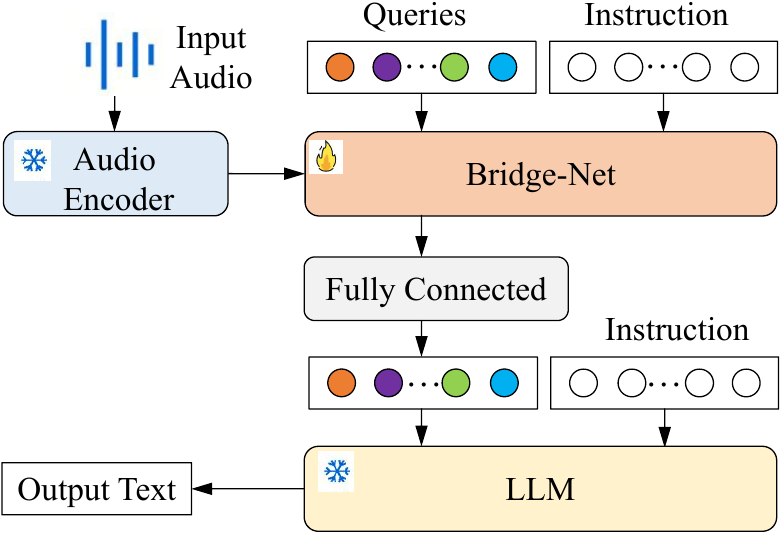}
  \caption{MINT's instruction tuning process.}
  \label{fig:adapt-pic}
  \vspace*{-\baselineskip}
\end{figure}

\subsection{Audio-language representation learning}
As shown in Figure 1, during the representation learning phase, Bridge-Net is interacted with the frozen audio encoder via cross-attention layers, facilitating pre-training with audio-caption pairs. The main objective of this training stage is to enable the learnable queries to extract audio representations that are most informative for the corresponding text input. Following BLIP-2 \cite{li2023blip}, our methodology incorporates three pre-training objectives: audio-language contrastive learning (ALC), audio-language matching (ALM), and audio-grounded text generation (ATG). As the audio transformer and text transformer share the self-attention layers in Bridge-Net, different attention mask strategies are applied to control the interaction between queries and text.

For the audio-language contrastive (ALC) learning, we aim to align audio and text representations by maximizing their mutual information \cite{mei2022metric}. This is achieved by contrasting the audio-language similarity of a positive pair against that of negative pairs. In Bridge-Net, the output query representation by the audio transformer is aligned with the text representation, which is the output embedding of the [CLS] token from the text transformer. As each query produces an output query embedding, the query representation contains multiple output query embeddings. We calculate the pairwise similarity between each query output and the [CLS] token, selecting the highest similarity as the audio-language similarity. Furthermore, we employ a self-attention mask strategy where the queries and text are not allowed to see each other to avoid information leak.

Audio-language matching (ALM) is designed to learn fine-grained alignment between audio and text embeddings. This is essentially a binary classifier where the model is required to determine if an audio-language pair is matched or unmatched \cite{li2022blip}. We feed each output query embedding into a two-class linear classifier to obtain a logit, and then average the logits across all queries as the output matching score. For ALM, we utilize a bi-directional self-attention mask that allows all queries and texts to attend to each other.

Audio-grounded text generation (ATG) trains the Bridge-Net to generate textual content conditioned on the input audio. Due to the architecture of Q-Former, direct interactions between the frozen audio encoder and the text tokens are not allowed. Therefore, the information required for text generation is initially extracted by the queries and then passed to the text tokens through self-attention layers. As a result, the queries are optimized to extract audio features that captures all the necessary information about the text as possible. To govern the interaction between queries and text tokens, we employ a causal self-attention mask \cite{dong2019unified} mechanism. This mechanism permits queries to attend to each other but not to the text tokens. On the other hand, each text token can attend to all queries and its preceding text tokens.

\subsection{Audio-to-language generative learning with Instruction Tuning}
During the generative training phase, we establish a connection between the Bridge-Net and a frozen LLM to leverage the LLM’s generative language capability. As illustrated in Figure 2, we employ a fully-connected (FC) layer to linearly project the output query embeddings to the same dimension as the text embeddings of the LLM. These projected query embeddings are then prepended to the input text embeddings, serving as soft audio prompts for the LLM. Since the Bridge-Net has been pre-trained to extract text-informative audio representation, it proficiently functions as an information bottleneck that feeds the most useful information to the LLM while filtering out irrelevant information. This approach alleviates the burden on the LLM to learn audio-language alignment, thus mitigating the issue of catastrophic forgetting.

What's more, to adapt to the task instructions \cite{dai2023instructblip} and generate audio representations that are highly relevant to the specific task, we boost Bridge-Net with an instruction-aware architecture. This enhanced mechanism incorporates the task instruction as additional text input. The instruction interacts with the query embeddings through self-attention layers of the Bridge-Net, promoting the extraction of audio features that are specific to the task instructions. As a result, the LLM receives audio information that is directly pertinent to the given instruction, thereby significantly enhancing the downstream performance on various tasks.

\begin{table}[h]\footnotesize
  \caption{The instruction templates. The $\{\}$ symbol indicates variable content. ``(t)" denotes the dataset used for training, while ``(e)" denotes the dataset applied for evaluation.}
  \vspace*{-\baselineskip}
  \centering
	\begin{tabular}{ccc}
		\hline Dataset & Input prompt & Output format \\\hline
		\makecell{AudioSet (t)} & This is a sound of &\{label\}\\\hline
		\makecell{VGGSound (t)}& This is a sound of &\{label\}\\\hline
		\makecell{Openmic (t+e)}& \makecell{Identify the instruments \\ in this segment of music: } &\{label\}\\\hline
		\makecell{FMALarge (t+e)}& The genre of this music is &\{label\}\\\hline
		\makecell{NSynth (t+e)}& \makecell{The most prominent \\instrument in this music is} &\{label\}\\\hline
		\makecell{FSD50K (t)}& This is a sound of  &\{label\}\\\hline
		\makecell{Music4All-key (t)}& The key of this music is  &\{label\}\\\hline
		\makecell{Music4All-genre (t)}& The genre of this music is &\{label\}\\\hline
		\makecell{GTZAN (e)}& The genre of this music is &\{label\}\\\hline
		\makecell{WavCaps (t)}&	Generate audio caption: &\{caption\}\\\hline
		\makecell{FreeSound (t)}& Generate audio caption: &\{caption\}\\\hline
		\makecell{Clotho (e)}& Generate audio caption: &\{caption\}\\\hline
	\end{tabular}
    \vspace*{-\baselineskip}
\end{table}	
\section{Experiments}
\subsection{Training datasets and templates}
Our training data is collected from multiple publicly available audio datasets. In total, we gather approximately 3 million audio-language pairs and categorize them into 12 templates, each based on different datasets. For the pre-training stage, we only use the WavCaps \cite{mei2023wavcaps} dataset for modality alignment training. For the instruction tuning stage, we leverage AudioSet \cite{gemmeke2017audio}, FSD50K \cite{fonseca2021fsd50k}, NSynth \cite{engel2017neural}, Music4all \cite{santana2020music4all}, FMA \cite{defferrard2016fma}, Openmic \cite{humphrey2018openmic}, Vggsound \cite{chen2020vggsound}, WavCaps \cite{mei2023wavcaps}, FreeSound \cite{fonseca2017freesound} for adapting to various audio tasks. 

Specifically, AudioSet \cite{gemmeke2017audio} consists of an expansive ontology of sound events and a vast collection of human-labeled 10-second sound clips drawn from YouTube videos. FSD50K \cite{fonseca2021fsd50k} is a sound dataset containing 51,000 audio clips. The NSynth dataset \cite{engel2017neural} is a large collection of over 300,000 musical notes, each annotated with a unique pitch, timbre, and envelope. Music4all\cite{santana2020music4all} offers a comprehensive suite for music understanding, featuring metadata for approximately 100,000 tracks. The Free Music Archive (FMA) dataset \cite{defferrard2016fma} encompasses a vast array of music tracks, totaling over 100,000 songs categorized by genre, subgenre, and additional metadata. The Open Music Instrument Classification dataset (OpenMIC)\cite{humphrey2018openmic} contains 20,000 audio snippets, each labeled with the presence or absence of one or more of 20 different musical instruments. VGGSound \cite{chen2020vggsound} is a large audio-visual dataset with over 200,000 clips from a diverse range of sources. Each clip includes both the audio track and the corresponding video frame, allowing for multimodal research. WavCaps \cite{mei2023wavcaps} is a recently introduced dataset that contains approximately 400k audio clips. It is designed to train and evaluate large-scale audio captioning models. FreeSound \cite{fonseca2017freesound} encompasses over 500k user-contributed sounds, spanning various categories and real-world scenarios. GTZAN  \cite{marchand2015gtzan} consists of 1,000 audio tracks, each 30 seconds long, evenly distributed across 10 different genres. Each genre is represented by 100 tracks, providing a balanced dataset for training and testing music genre classification models.

In the instruction tuning phase, our model is trained on a diverse collection of audio-language tasks formulated as instruction templates. This approach enhances the model's transferability and flexibility to handle various tasks during inference stage. The training datasets are modified to adapt to our method by constructing 10 audio-task templates. Each template is structured with an audio input, a associated input text prompt, and a corresponding text output. All the templates are in Table 1. 

\begin{table}\footnotesize
  \caption{For audio classification, we use Pengi as the baseline because of its strong performance on a wide range of downstream tasks. Higher is better for all numbers. “ZS” denotes the zero-shot setting.}
  \vspace*{-\baselineskip}
	\centering
	\begin{tabular}{c|c|cc|c|c}
		\hline	
		\multirow{2}{*}{Method}&\makecell{Nsynth}&\multicolumn{2}{c|}{OpenMic}&\makecell{FMA}&\makecell{GTZAN (ZS)}\\
        \cline{2-6}&ACC&PR&ROC&ACC&ACC\\\hline
		LAION & 29.49& 36.99 & 71.40& 45.38 & 28.31\\
		Pengi & 50.07& 41.34 & 72.88 & 46.22 & 35.25\\ 
		MINT & \textbf{68.26}& \textbf{46.72}& \textbf{77.91}& \textbf{49.28}& \textbf{49.66}\\\hline			
	\end{tabular}
 \vspace*{-\baselineskip}
\end{table}	
\subsection{Pre-training details}
For the frozen language model, we use FlanT5xl model \cite{chung2022scaling} as LLMs. We leverage data2vec \cite{baevski2022data2vec} as our audio encoder. The audio length ranges from 3s to 30s, while the max length of the text input is set to 30 for computational efficiency. We follow the training pipeline \cite{wu2023large} to train our models. We use 32 queries, where each query has a dimension of 768. The block number $N$ is 12. We use Adam Optimiser with a batch size of 64 on 8 A100 GPUs. We pre-train for 5 epochs in the first stage and 3 epochs in the second stage. We use a linear schedule with 2000 warmup steps and a base learning rate of 1e-4. 

\begin{table}
  \caption{Comparison with other text-to-audio retrieval methods on the Clotho dataset. Recall at rank k (R@k) metric indicates the percentage of relevant targets in the top-k ranked results. Higher is better for all numbers. The first four baseline models listed in the table are trained on the AudioCaps and Clotho datasets, while WavCaps and our MINT model are in a zero-shot setting.}
  \vspace*{-\baselineskip}
  \centering
  \begin{tabular}{c|ccc}
    \toprule
    \multirow{2}{*}{Method} & \multicolumn{3}{c}{Text-to-Audio} \\
    & \textbf{R@1} & \textbf{R@5} & \textbf{R@10}\\
    \midrule      
	MMT \cite{oncescu2021audio} & 6.7 & 21.6 & 33.2 \\
	ML-ACT \cite{mei2022metric} & 14.4 & 36.6 & 49.9 \\
	LAION \cite{wu2023large} & 16.4 & 39.0 & 51.0 \\
        Pengi \cite{deshmukh2023pengi} & 9.4 & 26.1 & 36.8 \\
        WavCaps \cite{mei2023wavcaps} & 16.5 & 38.8 & 50.9 \\
    \midrule  
	MINT & $\textbf{19.1}$ & $\textbf{39.3}$ & $\textbf{51.4}$ \\
  \bottomrule
\end{tabular}
\end{table}

\subsection{Benchmark}
MINT framework enables both close-ended and open-ended audio tasks simultaneously. We evaluate our method on both discriminative and generative audio-text downstream tasks. For discriminative tasks, our method is benchmarked on audio classification and text-audio retrieval tasks. For generative tasks, we evaluate on the audio captioning task. As pre-trained pure contrastive-based methods like CLAP, LAION are naturally unable to support the audio-to-language generation task without additional modules and fine-tuning, we compare our approach against Pengi and supervised trained models on the audio captioning task. 

\begin{table}\footnotesize
  \caption{Comparison with supervised audio captioning methods on the Clotho dataset. SPIDEr is the metric used to rank models in IEEE DCASE Challenge. Higher is better for all metrics. Note that only our MINT is in a zero-shot setting.}
  \vspace*{-\baselineskip}
  \centering
  \label{tab:freq}
  \begin{tabular}{cccc}
    \toprule
    Methods & SPIDEr \cite{spider2017} & ROUGE$_L$ \cite{rouge2004} & SPICE \cite{spice2016}\\
    \midrule
    BART-AAC \cite{gontier2021automated}  & 16.7 & 31.8 & 8.3\\
    ACT \cite{mei2021audio} & 20.9 & 35.1 & 10.5\\
    PT-AAC \cite{kim2023prefix} & 21.5 & 36.6 & 11.1\\
    Pengi \cite{deshmukh2023pengi} & 27.1 & 37.5 & 12.6\\
    \midrule  
    MINT & \textbf{28.2} & \textbf{36.2} & \textbf{12.8}\\
  \bottomrule
\end{tabular}
\vspace*{-\baselineskip}
\end{table}
\section{Results}
\subsection{Discriminative tasks}
\textbf{Audio classification task}. We evaluate audio-related classification tasks with a vocabulary ranking method \cite{wei2021finetuned}. Specifically, we prompt the LLM to generate results, but restrict its vocabulary to a list of classification candidates. Then, we calculate log-likelihood for each candidate and select the one with the highest score as the final prediction. As shown in Table 2, across all datasets and metrics, MINT consistently outperforms Pengi and LAION, demonstrating that MINT is a more robust and effective method for audio classification tasks. Notably, even in a zero-shot setting, MINT achieves an accuracy (ACC) of 49.66\%, marking a substantial enhancement of over 14 percentage points over Pengi. This underscores MINT's generalization ability to unseen data.

\begin{table}
  \caption{Influence of each part of the loss function in stage1.}
  \vspace*{-\baselineskip}
  \centering
  \begin{tabular}{ccc}
    \toprule
    Method & Nsynth & GTZAN (ZS)\\
    \midrule
    ALC & 61.31 & 41.27\\
    ALC+ATG & 66.72 & 43.34\\
    ALC+ALM & 62.76 & 46.13\\
    ALC+ALM+ATG & \textbf{68.26} & \textbf{49.66}\\
  \bottomrule
\end{tabular}
\vspace*{-\baselineskip}
\end{table}

\textbf{Audio-language retrieval task}. Since audio-language retrieval does not involve text generation, we directly evaluate the pre-trained stage1 model without the use of LLM, which are only trained on the Wavcaps dataset. Here, we follow the evaluation setting of LAION on the Clotho dataset. As shown in Table 3, despite not being trained on the Clotho dataset, MINT still outperforms those methods that have been trained on the Clotho dataset. Additionally, while both MINT and WavCaps are evaluated under a zero-shot setting, the distinct advantage of MINT is evident across all measured metrics, underscoring its robust capability in generalizing audio-language correlations.

\begin{table}
  \caption{Influence of MINT's different training stages.}
  \vspace*{-\baselineskip}
  \centering
  \begin{tabular}{ccc}
    \toprule
    Method & Nsynth & GTZAN (ZS)\\
    \midrule
    Stage1 & 55.91 & 41.87\\
    Stage1+2 & 68.26 & 49.66\\
  \bottomrule
\end{tabular}
\vspace*{-\baselineskip}
\end{table}
\subsection{Generative tasks}
\textbf{Audio captioning task}. Table 4 provides a performance comparison of various audio captioning methods on the Clotho dataset. Remarkably, even in a zero-shot setting, MINT exhibits competitive performance relative to other supervised approaches, especially in terms of the crucial SPIDEr metric. These results indicate that MINT holds substantial potential for zero-shot audio-to-text generation tasks, particularly in scenarios where audio-text data pairs may be limited or unavailable. 

\subsection{Ablation study}
In this section, we examine the impact of each component of the loss function in the initial stage and present the outcomes of various training stages.

\textbf{Influence of the loss function.} It can be seen in Table 5 that as a foundational loss function component, ALC yields an accuracy of 61.31 on Nsynth and 41.27 on GTZAN. By combining ALC with ATG, there's a noticeable increase in performance on the Nsynth dataset (66.72) compared to using ALC alone. However, the improvement is marginal on the GTZAN dataset (43.34). When ALM is combined with ALC, the performance reaches 62.76 on Nsynth and a commendable 46.13 on GTZAN (ZS). This suggests that ALM has a pronounced influence, especially on the GTZAN dataset. Merging all three components (ALC, ALM, and ATG) maximizes the performance, achieving 68.26 on Nsynth and 49.66 on GTZAN (ZS). This suggests that while each component contributes distinctively, their combined effect is synergistic, resulting in optimal outcomes.

\textbf{Results of different training stages.} As shown in Table 6, we further investigate the impact of the two training stages of MINT. In the pre-training stage, the absence of LLM necessitates the use of the audio-language embedding similarity ranking method, similar to the LAION series, for the classification task. In the instruction tuning stage, the aforementioned vocabulary ranking approach is adopted. It is evident that the optimal results are achieved upon the completion of both stages, providing strong evidence for the effectiveness of our approach. 

\section{Conclusions}
In this study, we introduce MINT, a novel ALP framework that enhances both audio-language understanding and generation tasks without requiring additional task-specific fine-tuning. By integrating multi-target learning and instruction tuning, MINT achieves superior performance across various audio-language tasks. Notably, MINT demonstrates robust generalizability when applied seamlessly to discriminative and generative downstream tasks in a zero-shot manner.

\bibliographystyle{IEEEtran}
\bibliography{mybib}

\begin{thebibliography}{10}
\providecommand{\url}[1]{#1}
\csname url@samestyle\endcsname
\providecommand{\newblock}{\relax}
\providecommand{\bibinfo}[2]{#2}
\providecommand{\BIBentrySTDinterwordspacing}{\spaceskip=0pt\relax}
\providecommand{\BIBentryALTinterwordstretchfactor}{4}
\providecommand{\BIBentryALTinterwordspacing}{\spaceskip=\fontdimen2\font plus
\BIBentryALTinterwordstretchfactor\fontdimen3\font minus \fontdimen4\font\relax}
\providecommand{\BIBforeignlanguage}[2]{{%
\expandafter\ifx\csname l@#1\endcsname\relax
\typeout{** WARNING: IEEEtran.bst: No hyphenation pattern has been}%
\typeout{** loaded for the language `#1'. Using the pattern for}%
\typeout{** the default language instead.}%
\else
\language=\csname l@#1\endcsname
\fi
#2}}
\providecommand{\BIBdecl}{\relax}
\BIBdecl

\bibitem{elizalde2023clap}
B.~Elizalde, S.~Deshmukh, M.~Al~Ismail, and H.~Wang, ``Clap learning audio concepts from natural language supervision,'' in \emph{ICASSP 2023-2023 IEEE International Conference on Acoustics, Speech and Signal Processing (ICASSP)}.\hskip 1em plus 0.5em minus 0.4em\relax IEEE, 2023, pp. 1--5.

\bibitem{wu2023large}
Y.~Wu, K.~Chen, T.~Zhang, Y.~Hui, T.~Berg-Kirkpatrick, and S.~Dubnov, ``Large-scale contrastive language-audio pretraining with feature fusion and keyword-to-caption augmentation,'' in \emph{ICASSP 2023-2023 IEEE International Conference on Acoustics, Speech and Signal Processing (ICASSP)}.\hskip 1em plus 0.5em minus 0.4em\relax IEEE, 2023, pp. 1--5.

\bibitem{deshmukh2023pengi}
S.~Deshmukh, B.~Elizalde, R.~Singh, and H.~Wang, ``Pengi: An audio language model for audio tasks,'' \emph{Advances in Neural Information Processing Systems}, vol.~36, 2024.

\bibitem{kenton2019bert}
J.~D. M.-W.~C. Kenton and L.~K. Toutanova, ``Bert: Pre-training of deep bidirectional transformers for language understanding,'' in \emph{Proceedings of naacL-HLT}, 2019, pp. 4171--4186.

\bibitem{cheng2023ml}
X.~Cheng, B.~Cao, Q.~Ye, Z.~Zhu, H.~Li, and Y.~Zou, ``Ml-lmcl: Mutual learning and large-margin contrastive learning for improving asr robustness in spoken language understanding,'' in \emph{Findings of the Association for Computational Linguistics: ACL 2023}, 2023, pp. 6492--6505.

\bibitem{cheng2023mrrl}
X.~Cheng, Z.~Zhu, B.~Cao, Q.~Ye, and Y.~Zou, ``Mrrl: Modifying the reference via reinforcement learning for non-autoregressive joint multiple intent detection and slot filling,'' in \emph{The 2023 Conference on Empirical Methods in Natural Language Processing}, 2023, pp. 10\,495--10\,505.

\bibitem{23d_interspeech}
Y.~Xin, X.~Peng, and Y.~Lu, ``Masked audio modeling with clap and multi-objective learning,'' in \emph{Proc. INTERSPEECH 2023}, 2023, pp. 2763--2767.

\bibitem{2023pooling}
Y.~Xin, D.~Yang, and Y.~Zou, ``Improving text-audio retrieval by text-aware attention pooling and prior matrix revised loss,'' in \emph{ICASSP 2023-2023 IEEE International Conference on Acoustics, Speech and Signal Processing (ICASSP)}.\hskip 1em plus 0.5em minus 0.4em\relax IEEE, 2023, pp. 1--5.

\bibitem{2023cooperative}
Y.~Xin, B.~Wang, and L.~Shang, ``Cooperative game modeling with weighted token-level alignment for audio-text retrieval,'' \emph{IEEE Signal Processing Letters}, 2023.

\bibitem{gontier2021automated}
F.~Gontier, R.~Serizel, and C.~Cerisara, ``Automated audio captioning by fine-tuning bart with audioset tags,'' in \emph{DCASE 2021-6th Workshop on Detection and Classification of Acoustic Scenes and Events}, 2021.

\bibitem{mei2021audio}
X.~Mei, X.~Liu, Q.~Huang, M.~D. Plumbley, and W.~Wang, ``Audio captioning transformer,'' in \emph{Proceedings of the 6th Detection and Classification of Acoustic Scenes and Events 2021 Workshop (DCASE2021)}, Barcelona, Spain, November 2021, pp. 211--215.

\bibitem{li2023blip}
J.~Li, D.~Li, S.~Savarese, and S.~Hoi, ``Blip-2: bootstrapping language-image pre-training with frozen image encoders and large language models,'' in \emph{Proceedings of the 40th International Conference on Machine Learning}, ser. ICML'23, 2023.

\bibitem{mei2022metric}
X.~Mei, X.~Liu, J.~Sun, M.~Plumbley, and W.~Wang, ``{On Metric Learning for Audio-Text Cross-Modal Retrieval},'' in \emph{Proc. Interspeech 2022}, 2022, pp. 4142--4146.

\bibitem{li2022blip}
J.~Li, D.~Li, C.~Xiong, and S.~Hoi, ``Blip: Bootstrapping language-image pre-training for unified vision-language understanding and generation,'' in \emph{International Conference on Machine Learning}.\hskip 1em plus 0.5em minus 0.4em\relax PMLR, 2022, pp. 12\,888--12\,900.

\bibitem{dong2019unified}
L.~Dong, N.~Yang, W.~Wang, F.~Wei, X.~Liu, Y.~Wang, J.~Gao, M.~Zhou, and H.-W. Hon, ``Unified language model pre-training for natural language understanding and generation,'' \emph{Advances in neural information processing systems}, vol.~32, 2019.

\bibitem{dai2023instructblip}
W.~Dai, J.~Li, D.~Li, A.~Tiong, J.~Zhao, W.~Wang, B.~Li, P.~Fung, and S.~Hoi, ``Instruct{BLIP}: Towards general-purpose vision-language models with instruction tuning,'' in \emph{Thirty-seventh Conference on Neural Information Processing Systems}, 2023.

\bibitem{mei2023wavcaps}
X.~Mei, C.~Meng, H.~Liu, Q.~Kong, T.~Ko, C.~Zhao, M.~D. Plumbley, Y.~Zou, and W.~Wang, ``Wavcaps: A chatgpt-assisted weakly-labelled audio captioning dataset for audio-language multimodal research,'' \emph{arXiv preprint arXiv:2303.17395}, 2023.

\bibitem{gemmeke2017audio}
J.~F. Gemmeke, D.~P. Ellis, D.~Freedman, A.~Jansen, W.~Lawrence, R.~C. Moore, M.~Plakal, and M.~Ritter, ``Audio set: An ontology and human-labeled dataset for audio events,'' in \emph{2017 IEEE international conference on acoustics, speech and signal processing (ICASSP)}.\hskip 1em plus 0.5em minus 0.4em\relax IEEE, 2017, pp. 776--780.

\bibitem{fonseca2021fsd50k}
E.~Fonseca, X.~Favory, J.~Pons, F.~Font, and X.~Serra, ``Fsd50k: an open dataset of human-labeled sound events,'' \emph{IEEE/ACM Transactions on Audio, Speech, and Language Processing}, vol.~30, pp. 829--852, 2021.

\bibitem{engel2017neural}
J.~Engel, C.~Resnick, A.~Roberts, S.~Dieleman, M.~Norouzi, D.~Eck, and K.~Simonyan, ``Neural audio synthesis of musical notes with wavenet autoencoders,'' in \emph{International Conference on Machine Learning}.\hskip 1em plus 0.5em minus 0.4em\relax PMLR, 2017, pp. 1068--1077.

\bibitem{santana2020music4all}
I.~A.~P. Santana, F.~Pinhelli, J.~Donini, L.~Catharin, R.~B. Mangolin, V.~D. Feltrim, M.~A. Domingues \emph{et~al.}, ``Music4all: A new music database and its applications,'' in \emph{2020 International Conference on Systems, Signals and Image Processing (IWSSIP)}.\hskip 1em plus 0.5em minus 0.4em\relax IEEE, 2020, pp. 399--404.

\bibitem{defferrard2016fma}
M.~Defferrard, K.~Benzi, P.~Vandergheynst, and X.~Bresson, ``Fma: A dataset for music analysis,'' in \emph{18th International Society for Music Information Retrieval Conference}, no. CONF, 2017.

\bibitem{humphrey2018openmic}
E.~Humphrey, S.~Durand, and B.~McFee, ``Openmic-2018: An open data-set for multiple instrument recognition.'' in \emph{ISMIR}, 2018, pp. 438--444.

\bibitem{chen2020vggsound}
H.~Chen, W.~Xie, A.~Vedaldi, and A.~Zisserman, ``Vggsound: A large-scale audio-visual dataset,'' in \emph{ICASSP 2020-2020 IEEE International Conference on Acoustics, Speech and Signal Processing (ICASSP)}.\hskip 1em plus 0.5em minus 0.4em\relax IEEE, 2020, pp. 721--725.

\bibitem{fonseca2017freesound}
E.~Fonseca, J.~Pons~Puig, X.~Favory, F.~Font~Corbera, D.~Bogdanov, A.~Ferraro, S.~Oramas, A.~Porter, and X.~Serra, ``Freesound datasets: a platform for the creation of open audio datasets,'' in \emph{ISMIR}.\hskip 1em plus 0.5em minus 0.4em\relax International Society for Music Information Retrieval (ISMIR), 2017.

\bibitem{marchand2015gtzan}
U.~Marchand, Q.~Fresnel, and G.~Peeters, ``Gtzan-rhythm: Extending the gtzan test-set with beat, downbeat and swing annotations,'' 2015.

\bibitem{chung2022scaling}
H.~W. Chung, L.~Hou, S.~Longpre, B.~Zoph, Y.~Tay, W.~Fedus, E.~Li, X.~Wang, M.~Dehghani, S.~Brahma \emph{et~al.}, ``Scaling instruction-finetuned language models,'' \emph{arXiv preprint arXiv:2210.11416}, 2022.

\bibitem{baevski2022data2vec}
A.~Baevski, W.-N. Hsu, Q.~Xu, A.~Babu, J.~Gu, and M.~Auli, ``Data2vec: A general framework for self-supervised learning in speech, vision and language,'' in \emph{International Conference on Machine Learning}.\hskip 1em plus 0.5em minus 0.4em\relax PMLR, 2022, pp. 1298--1312.

\bibitem{oncescu2021audio}
A.~S. Koepke, A.-M. Oncescu, J.~Henriques, Z.~Akata, and S.~Albanie, ``Audio retrieval with natural language queries: A benchmark study,'' \emph{IEEE Transactions on Multimedia}, 2022.

\bibitem{spider2017}
C.~Liu, R.~Lowe, I.~V. Serban, M.~Noseworthy, L.~Charlin, and J.~Pineau, ``Spider: A robust metric for evaluating content preservation in neural machine translation,'' in \emph{Proceedings of the 55th Annual Meeting of the Association for Computational Linguistics (Volume 2: Short Papers)}, 2017, pp. 148--153.

\bibitem{rouge2004}
C.-Y. Lin, ``Rouge: A package for automatic evaluation of summaries,'' in \emph{Text Summarization Branches Out}.\hskip 1em plus 0.5em minus 0.4em\relax Barcelona, Spain, 2004.

\bibitem{spice2016}
P.~Anderson, B.~Fernando, M.~Johnson, and S.~Gould, ``Spice: Semantic propositional image caption evaluation,'' in \emph{European Conference on Computer Vision}.\hskip 1em plus 0.5em minus 0.4em\relax Springer, 2016, pp. 382--398.

\bibitem{kim2023prefix}
M.~Kim, K.~Sung-Bin, and T.-H. Oh, ``Prefix tuning for automated audio captioning,'' in \emph{ICASSP 2023-2023 IEEE International Conference on Acoustics, Speech and Signal Processing (ICASSP)}.\hskip 1em plus 0.5em minus 0.4em\relax IEEE, 2023, pp. 1--5.

\bibitem{wei2021finetuned}
J.~Wei, M.~Bosma, V.~Zhao, K.~Guu, A.~W. Yu, B.~Lester, N.~Du, A.~M. Dai, and Q.~V. Le, ``Finetuned language models are zero-shot learners,'' in \emph{International Conference on Learning Representations}, 2022.

\end{thebibliography}

\end{document}